\documentclass[aps,prl,10pt,twocolumn,preprintnumbers,superscriptaddress,nofootinbib,letterpaper,longbibliography]{revtex4-2}

\usepackage{hyperref}
\usepackage{graphicx}
\usepackage{amsfonts,amsmath,amssymb,bm}
\usepackage{array}
\usepackage{xcolor}
\usepackage{comment}
\usepackage{soul}
\usepackage[utf8]{inputenc}
\usepackage{tikz}
\usepackage{multirow}
\usepackage{boldline}
\usepackage{svg}
\usepackage{todonotes}
\usepackage{hyperref}
\usepackage{slashed}
\usepackage{orcidlink}
\hypersetup{
  colorlinks  = true,
  urlcolor    = blue,
  linkcolor   = blue,
  citecolor   = red
}

\newcolumntype{P}[1]{>{\centering\arraybackslash}p{#1}}

\def\Fermilab{Theory Division, Fermilab, P.O. Box 500, Batavia, IL 60510, USA}

\graphicspath{{figs/}}

\begin{document}


\title{Revisiting Gauge Ambiguities for DUNE Precision}

\author{Joshua Isaacson\,\orcidlink{0000-0001-6164-1707}}
\email{isaacson@fnal.gov}
\affiliation{\Fermilab}

\date{\today}


\begin{abstract}
The de Forest prescription for handling off-shell initial states in the impulse approximation for lepton-nucleus scattering breaks gauge invariance. We discuss existing methods to address this problem and handle the form factor scale ambiguity.
We demonstrate that the irreducible differences between the prescriptions are significant compared to the precision expected of next-generation accelerator neutrino experiments.
A novel approach directly using the off-shell currents is proposed as a systematically improvable alternative.
\end{abstract}

\preprint{FERMILAB-PUB-25-0059-T}

\maketitle


Next-generation accelerator neutrino experiments will provide an unprecedented experimental test of the nature of neutrino oscillations~\cite{MicroBooNE:2015bmn,JUNO:2015zny,Hyper-Kamiokande:2018ofw,DUNE:2020ypp}.
For the first time, the measurements will be limited by systematic uncertainties.
It will be of vital importance to completely account for all possible contributions to these uncertainties.

A significant systematic uncertainty in current and future neutrino experiments arises from the theoretical calculations involved in neutrino-nucleus interactions~\cite{NOvA:2019cyt}.
This challenge is particularly acute for experiments like the Deep Underground Neutrino Experiment (DUNE), which aims to make precision measurements of neutrino oscillation parameters, determine the mass hierarchy, measure the extent of CP violation in the lepton sector, and probe weak coupling light mass new physics models~\cite{DUNE:2020fgq}.
To achieve these ambitious goals, DUNE will require unprecedented precision in neutrino-nucleus interaction modeling, with uncertainties constrained to the level of a few percent~\cite{DUNE:2020ypp}.

Accurately modeling these interactions presents various challenges~\cite{Ruso:2022qes}.
For example, form factors associated with the structure of the proton can be either calculated from first principles using lattice QCD or fitted to experimental data~\cite{Bhattacharya:2011ah, Meyer:2016oeg, Davoudi:2020ngi, Borah:2020gte, Meyer:2022mix}.
To achieve percent-level uncertainties, many-body correlation effects within factorization schemes such as the impulse approximation need to be considered~\cite{Benhar:2006wy, Martini:2009uj, Martini:2010ex, Amaro:2010sd,Gran:2013kda, Benhar:2015ula}.
Final-state interactions are extremely complex and require the computation of non-perturbative single and many-nucleon effects~\cite{Lovato:2016gkq, Lovato:2020kba, Sobczyk:2021dwm}, approximated using intranuclear cascades~\cite{Serber:1947zza,Metropolis:1958sb,Bertini:1963zzc,Cugnon:1980zz,Boudard:2002yn,Hayato:2002sd,Casper:2002sd,Battistoni:2013tra,Battistoni:2015epi,Golan:2012wx,Andreopoulos:2009rq,Isaacson:2020wlx} or transport equations~\cite{KadanoffBaym:1962, Botermans:1990qi, Cassing:1990dr,Teis:1996kx, Buss:2011mx,Mosel:2019vhx}.

In this work, we focus on a critical aspect of these calculations: the treatment of bound initial state nucleons within the plane-wave impulse approximation (PWIA), which may be treated as off-shell particles. While here we will focus on only quasielastic scattering, the same issues arise in the handling of any process that contains off-shell initial state nucleons. The PWIA framework, widely employed in both theoretical predictions~\cite{DEFOREST1967365,DeForest:1983ahx} and neutrino-nucleus event generators~\cite{Andreopoulos:2009rq,Buss:2011mx,Golan:2012rfa,MINERvA:2013zvz,Hayato:2021heg,Isaacson:2022cwh,MicroBooNE:2024zkh}, represents the interaction as occurring with individual nucleons treated as quasi-free particles.
This approximation assumes that both initial and final state nucleons can be described using plane-wave wavefunctions, effectively treating them as free particles modified by nuclear effects.

While the PWIA has proven remarkably successful for high-energy interactions, where the quasi-free nucleon picture becomes increasingly valid, the treatment of off-shell effects has never been systematically studied to determine the associated uncertainties.
These effects become particularly relevant when considering that bound nucleons in the nucleus have an effective mass that deviates from the free-particle mass, potentially affecting both cross sections and kinematic distributions in neutrino-nucleus interactions.

Our investigation systematically examines different approaches to handling off-shell effects within the PWIA framework, quantifying their impact on predicted observables and associated uncertainties.
This analysis is crucial for developing a complete error budget for neutrino-nucleus interaction calculations and ultimately achieving the precision required by next-generation experiments.

The fully-differential cross section for lepton-nucleus scattering processes is given by (See Sec. II of Ref.~\cite{Isaacson:2022cwh} for additional details)
\begin{align}
    d\sigma =& \left(\frac{1}{|v_A-v_\ell|}\frac{1}{4E_A^{\rm in} E_\ell^{\rm in}}\right) L_{\mu\nu}W^{\mu\nu}\frac{1}{P^2} \nonumber\\
    &\times\prod_f\frac{d^3p_f}{(2\pi)^3}(2\pi)^4 \delta^4\Big(k_A+k_\ell - \sum_f p_f\Big)\,,
    \label{eq:xsec_general}
\end{align}
where $L_{\mu\nu}$ is the leptonic tensor, $W^{\mu\nu}$ is the hadronic tensor, and the initial- and final-state momenta are defined as
\begin{align}\label{eq:notation}
  &k_\ell^\mu=(E_\ell^{\rm in},{\bf k}) \quad\text{incoming}~\ell,\\
  &k_A^\mu=(E_A^{\rm in},{\bf k}_A)\quad\text{incoming}~A,\\
  &p_f^\mu=(E_f,{\bf p}_f)\quad\text{outgoing~particle~}f,\label{eq:notation-last}
\end{align}
where the index $f$ refers to all the possible hadronic and leptonic final state particles.
The leptonic tensor involves no nuclear physics and can be calculated in an automated manner~\cite{Isaacson:2021xty}.
On the other hand, the hadronic tensor involves all the underlying nuclear physics, and takes the general form
\begin{equation}\label{eq:had_tensor}
    W^{\mu\nu} = \langle \Psi_0 | J^{\dagger\mu} (q) | \Psi_f \rangle \langle \Psi_f | J^\nu (q) | \Psi_0 \rangle\,.
\end{equation}
Here, the sates $\Psi_0$ and $\Psi_f$ denote the full hadronic initial and final states, respectively (i.e. including all nucleons), $J^{\mu}$ is an arbitrary Electroweak current, and $q=(\omega,\bf{q})$ is the momentum transfer given by the difference between the sum of final state leptons and the initial state lepton momenta.

Applying the PWIA, inserting a complete set of states twice ($\int {\rm d}k\lvert \Psi_f^{A-1}\otimes k\rangle\langle \Psi_f^{A-1}\otimes k\rvert$) and retaining only the one-body current contributions to Eq.~\eqref{eq:had_tensor}, one obtains
\begin{align}
W^{\mu\nu}({\bf q}, \omega)&=\!\! \sum_{h \in \{p,n\}} \int  \frac{d^3k_h}{(2\pi)^3}\frac{1}{2E_h^{\rm in}} dE^\prime S_h({\bf k}_h,E_\prime) \nonumber\\
 & \times \langle k| {j_{1b}^\mu}^\dagger|p \rangle \langle p| j_{1b}^\nu| k \rangle (2\pi)^4 \delta^3({\bf k}_h^{\rm}+{\bf q}-{\bf p}_h)\nonumber\\ 
 & \times \delta(\omega-E^\prime+m_N-E_h)\,,
 \label{eq:hadronic_tensor}
\end{align}
where $\lvert \Psi_f^{A-1}\otimes k\rangle$ is the dyadic product of a single-nucleon state with momentum $k$ and the state of the remaining $A-1$ nucleons.
In the above equation, $k_h=(E_h^{\rm in}, \textbf{k}_h)$ and $p_h=(E_h, \textbf{p}_h)$ are the momenta of the initial and final state nucleons, respectively.
The spectral function ($S_h(\textbf{k}_h, E^\prime)$) represents the probability distribution associated with removing a ``hole'' nucleon with three-momentum $\bf{k}_h$ from the target nucleus and leaving the residual system with an excitation energy $E^\prime$. It is defined as~\cite{Benhar:1993ja}
\begin{align}\label{eq:SF}
S_h({\bf k}_h,& E^\prime, s_i, s_j) = \\ 
\sum_{f_{A-1}} &\langle \Psi_0|k_{s_i} \otimes \Psi_f^{A-1}\rangle \langle k_{s_j} \otimes \Psi_f^{A-1}|\Psi_0\rangle \nonumber \\ &\delta(E^\prime+E_0^A-E_f^{A-1})\,.\nonumber
\end{align}
The sum in the above equation is over all possible final states of the remaining $A-1$ spectator nucleons, while $s_{i,j}$ represent the spins of the initial state nucleons. Traditionally, the spin contribution is neglected in the PWIA since it only contributes if there is parity violation~\cite{DeForest:1983ahx}.

In order to evaluate the hadronic tensor, one needs to compute the one-body current given as $\langle k| j_{1b}^\mu| p \rangle$ for an off-shell initial state nucleon~\footnote{Here, we will assume that the outgoing nucleon is on-shell, see Ref.~\cite{Naus:1990em} for the more general case}.
The most general form of this current is given by~\cite{Naus:1987kv}
\begin{align}
    j_{1b}^\mu = &\left[ \left( C_1 \gamma^{\mu}f_1^{(+)}\left(W,q^2\right) + C_2 \frac{i\sigma^{\mu\nu}}{2M}q_\nu f_2^{(+)}\left(W,q^2\right)\right.\right. \nonumber \\
    &\left.\left. +C_3 q^\mu f_3^{(+)}\left(W,q^2\right)+C_A\gamma^{\mu}\gamma_5 f_a^{(+)}\left(W,q^2\right)\right.\right. \nonumber \\
    &\left.\left.+C_{AP} \frac{q^\mu}{M} \gamma_5 f_{ap}^{(+)}\left(W,q^2\right) \right)\Lambda_+ \right. \nonumber \\
    &\left. + \left(C_1\gamma^{\mu}f_1^{(-)}\left(W,q^2\right)+C_2\frac{i\sigma^{\mu\nu}}{2M}q_\nu f_2^{(-)}\left(W,q^2\right)\right.\right. \nonumber \\
    &\left.\left. +C_3 q^\mu f_3^{(-)}\left(W,q^2\right)+\gamma^{\mu}\gamma_5 C_A f_a^{(-)}\left(W,q^2\right)\right.\right.\nonumber \\
    & \left.\left.+C_{AP} \frac{q^\mu}{M} \gamma_5 f_{ap}^{(-)}\left(W,q^2\right) \right)\Lambda_{-}\right]\,,
    \label{eq:offshell_current}
\end{align}
where $C_i$ is the nucleon coupling to the gauge boson which may depend on the isospin of the nucleons, $f_{i}^{(\pm)}$ are structure functions that depend on both the momentum transfer ($q^2$) and the off-shellness of the nucleon ($p^2$), and the projection operators $\Lambda_{\pm}$ are defined for general four-momenta $p$ as
\begin{equation}
    \Lambda_{\pm}(p) = \frac{W\pm \slashed{p}}{2 W}.
\end{equation}
In the above equation $W$ is defined as $\sqrt{p^2}$.
The projection operators satisfy the identities
\begin{align}
   \Lambda_{\pm}^2 &= \Lambda_{\pm}\,, \\
   \Lambda_{+}\Lambda_{-} &= \Lambda_{-}\Lambda_{+} = 0\,, \\
   \Lambda_{+} &+ \Lambda_{-} = \mathbf{I}\,, \\
   \Lambda_{+} &- \Lambda_{-} = \frac{\slashed{p}}{W}\,.
\end{align}
In the on-shell case ($W=M$), the projection operators are those for the positive and negative energy states. 
Taking the on-shell limit of Eq.~\eqref{eq:offshell_current}, one obtains the traditional formula used for describing nucleon-boson interactions, namely
\begin{equation}
    \begin{split}
    j_{1b}^\mu = C_1\gamma^\mu F_1(q^2) + C_2\frac{i\sigma^{\mu\nu}q_\nu}{2M}F_2(q^2) \\
    + C_A \gamma^\mu \gamma_5 F_A(q^2) +C_{AP} \gamma_5 \frac{q^\mu}{M} F_{AP}(q^2)\,,
    \end{split}
    \label{eq:on-shell_current}
\end{equation}
where $F_1$, $F_2$, $F_A$, and $F_{AP}$ are the Dirac, Pauli, axial, and pseudo-axial form factors, respectively.
The relationship between the on-shell form factors and the off-shell ones is given by $F_i(q^2) = f_i(q^2, M^2)$.

Since lattice calculations and experimental extractions of the form factors are performed with the assumption of an on-shell nucleon, many event generators chose to only consider the on-shell form factors in their calculation and treat the initial state wave function as purely on-shell.
However, the spectral function inherently requires that the initial state nucleons are off-shell.
This is typically addressed through the de Forest prescription by shifting the energy of the gauge boson as~\cite{DeForest:1983ahx}
\begin{equation}
   \tilde{\omega} = \omega + E_h^{\rm in} - \bar{E},
\end{equation}
where $\bar{E}$ is the energy for an on-shell nucleon with momentum $p$ and $E_h^{\rm in}$ is again the energy of the initial state off-shell nucleon.
The shift in the energy now results in the violation of the Ward Identity in Eq.~\eqref{eq:on-shell_current}, and thus currents are not conserved.
To restore current conservation, Ref.~\cite{Kelly:1997zza} points out three possible prescriptions that are associated with the gauge that gives the same results without modifying the currents.
The first is the Coulomb gauge, originally proposed by de Forest~\cite{DeForest:1983ahx}
\begin{equation}
    j_{1b}^q \rightarrow \frac{\omega}{q} j_{1b}^0\,,
\end{equation}
where the $q$-component corresponds to the longitudinal component of the current, and $0$ corresponds to the energy component.
The second prescription is associated with the Weyl gauge and is defined as
\begin{equation}
    j_{1b}^0 \rightarrow \frac{q}{\omega} j_{1b}^q\,.
    \label{eq:weyl}
\end{equation}
Finally, the Landau gauge prescription is given as
\begin{equation}
j_{1b}^\mu \rightarrow j_{1b}^\mu + \frac{j_{1b}\cdot q}{Q^2}q^\mu\,,
\end{equation}
which was first used in Ref.~\cite{Mougey:1976sc}.
It is important to note that in the Landau prescription, the modification does not contribute to changing the rate for the case of electron scattering or neutral current neutrino scattering, and thus is the same as using the original non-conserved current. 
This is due to the fact that the contribution from the $q^\mu$ term disappears when the masses of the initial and final state lepton are the same.
However, the Landau prescription has an effect in the case of charged current neutrino scattering due to the difference in mass between the neutrino and the charged lepton.

In addition to the method of gauge restoration, the choice to shift the energy of the gauge boson also introduces an ambiguity in the scale used in the form factors.
There are two possible choices that are reasonable to make. The first is to use the original gauge boson momentum ($Q = (\omega, \vec{q})$) and the second choice is to use the shifted gauge boson momentum ($\tilde{Q} = (\tilde{\omega}, \vec{q})$). 

\begin{figure}[!ht]
    \centering
    \includegraphics[width=0.92\linewidth]{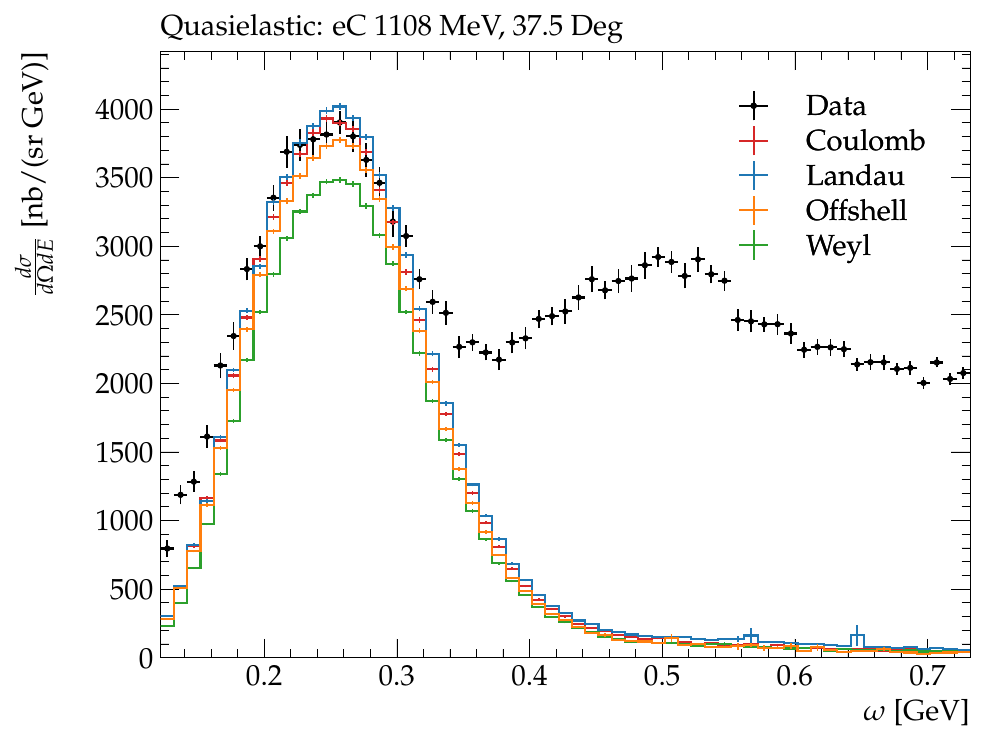} 
    \\
    \includegraphics[width=0.92\linewidth]{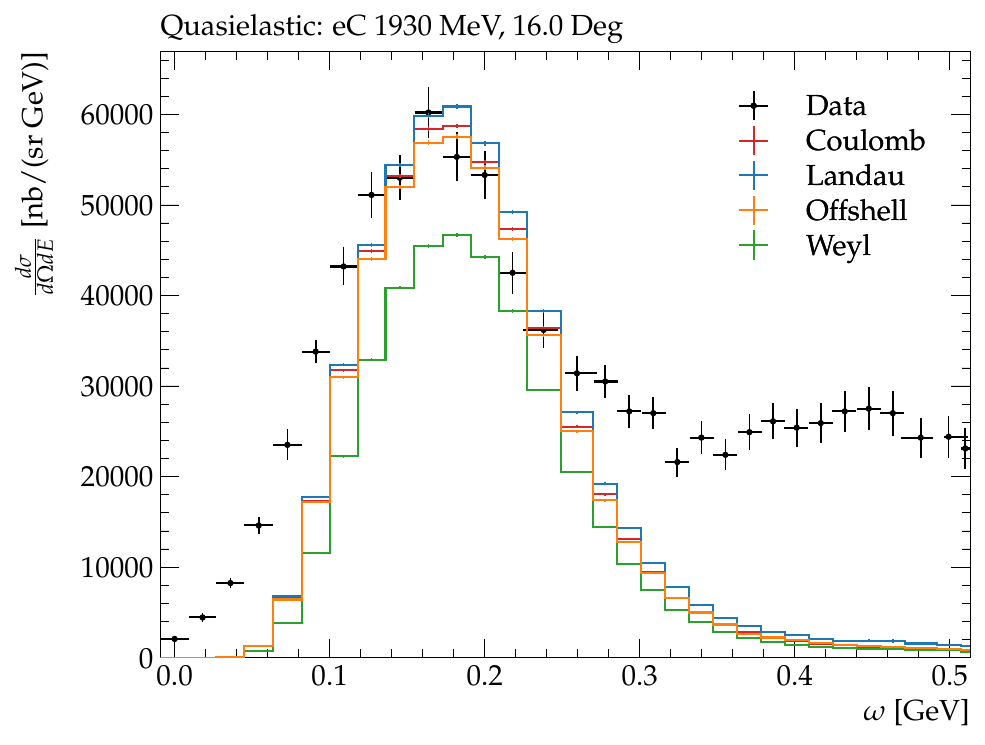}
    \\
    \includegraphics[width=0.92\linewidth]{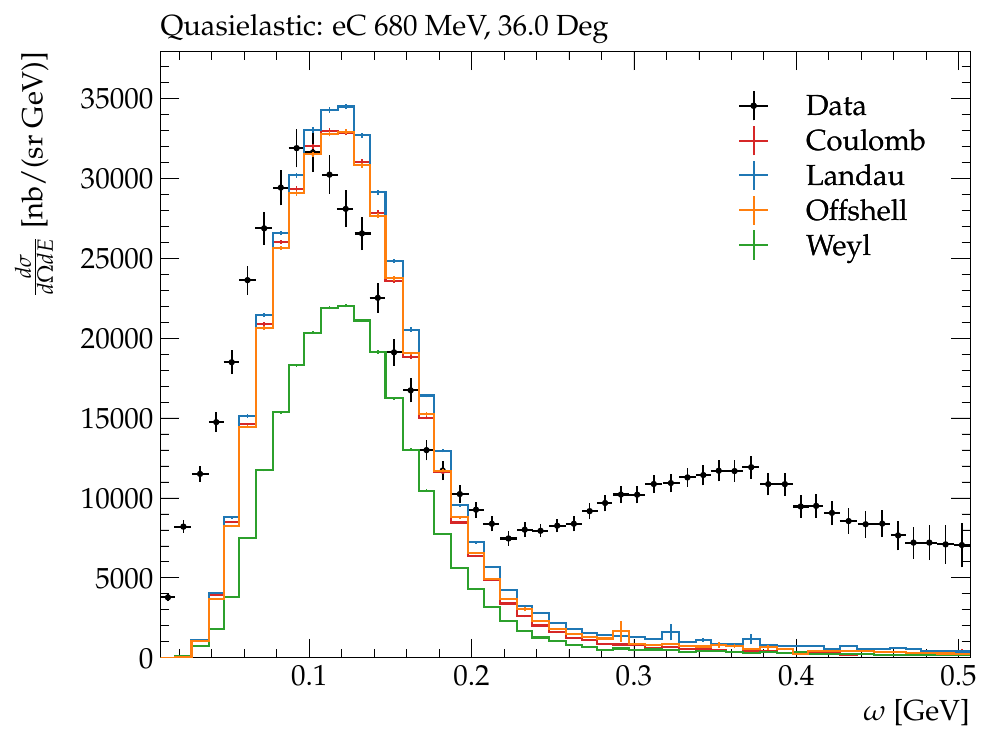}
    \caption{A comparison of different quasielastic calculations for maintaining current conservation for electron-carbon scattering at various outgoing electron angles and beam energies. The Coulomb prescription is shown in red, the Landau prescription is shown in blue, the Weyl prescription is shown in green, and the off-shell prescription is shown in orange. A detailed description of each prescription is given in the text. The experimental data comes from Ref.~\cite{Sealock:1989nx} (top), Ref.~\cite{Bagdasaryan:1988hp} (middle), and Ref.~\cite{Barreau:1983ht} (bottom). Note that the theory calculations neglect contributions from meson-exchange currents and resonance contributions that arise at large $\omega$.}
    \label{fig:electron_scattering}
\end{figure}

\begin{figure}[!ht]
    \centering
    \includegraphics[width=\linewidth]{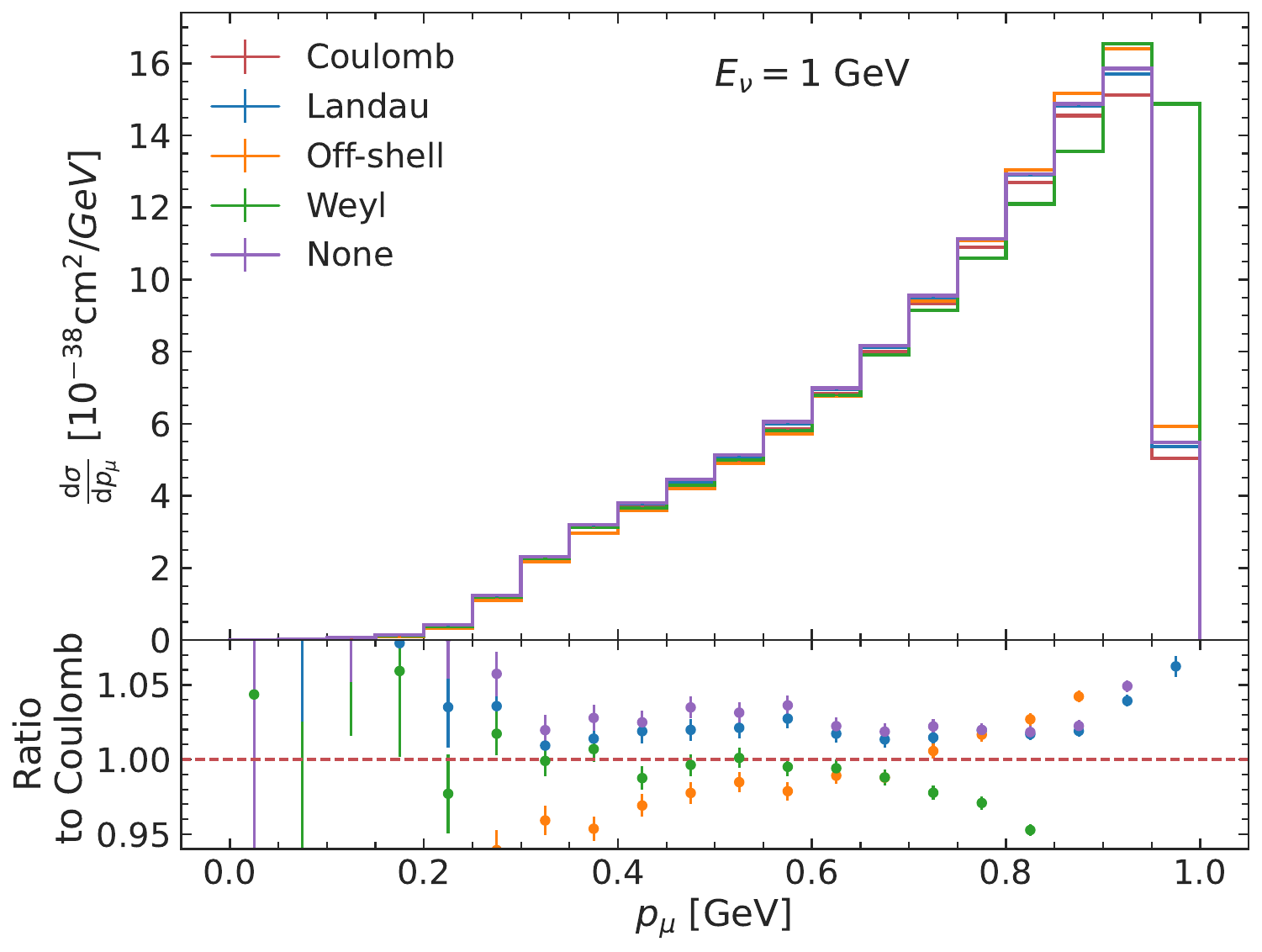} 
    \\
    \includegraphics[width=\linewidth]{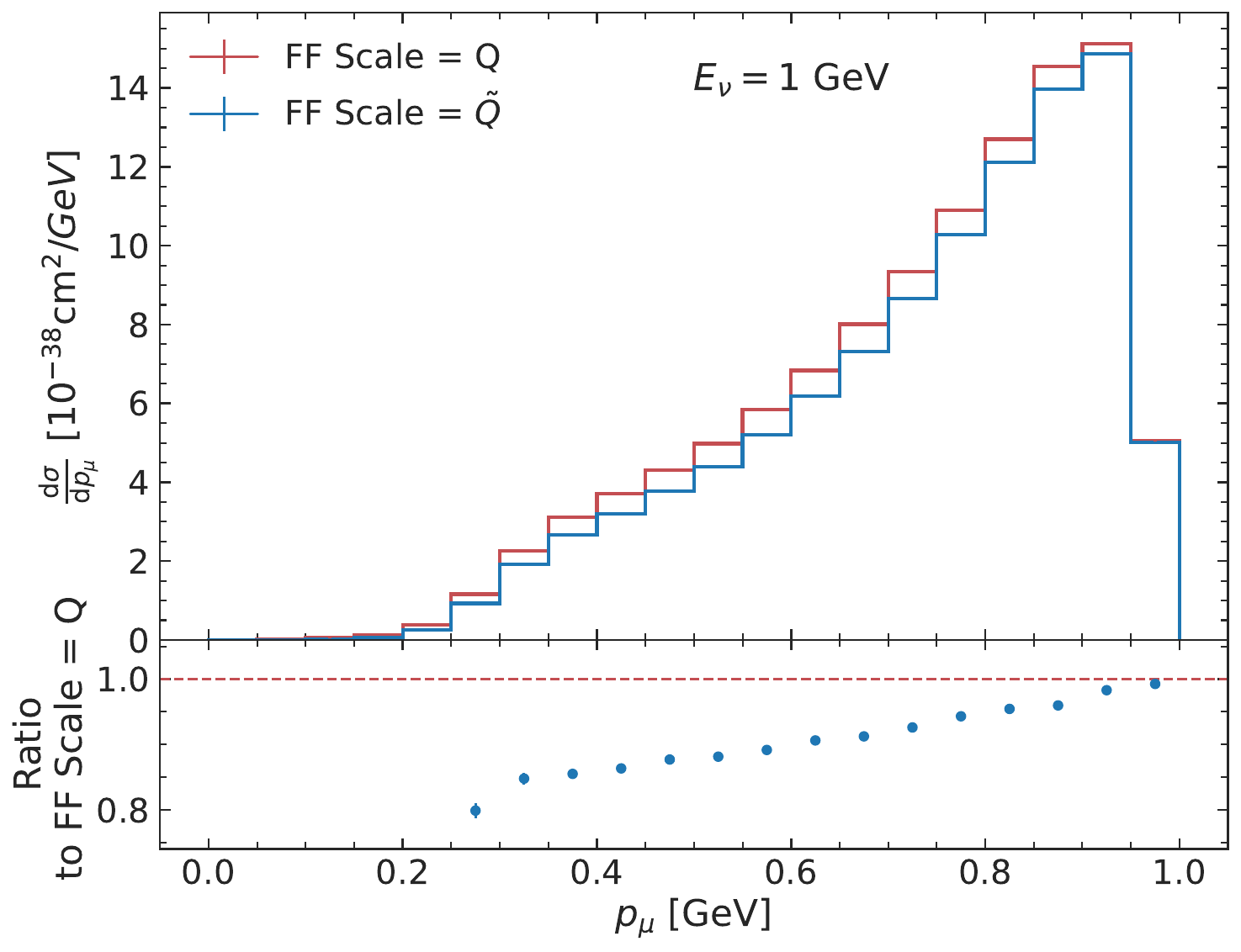}
    \caption{A comparison of different calculations for maintaining current conservation for neutrino-carbon scattering for a 1 GeV neutrino beam. The top plot shows the outgoing muon momentum for different prescriptions for gauge restoration, with Coulomb in red, Landau in blue, the Off-shell in orange, and Weyl in green.
    The bottom plot shows the different choices for the scale of the form factors, depending on if the original gauge boson momentum is used or the shifted momentum.
    A detailed description of each prescription and form factor choice is given in the text. 
    }
    \label{fig:neutrino_scattering_mom}
\end{figure}

\begin{figure}[!ht]
    \includegraphics[width=\linewidth]{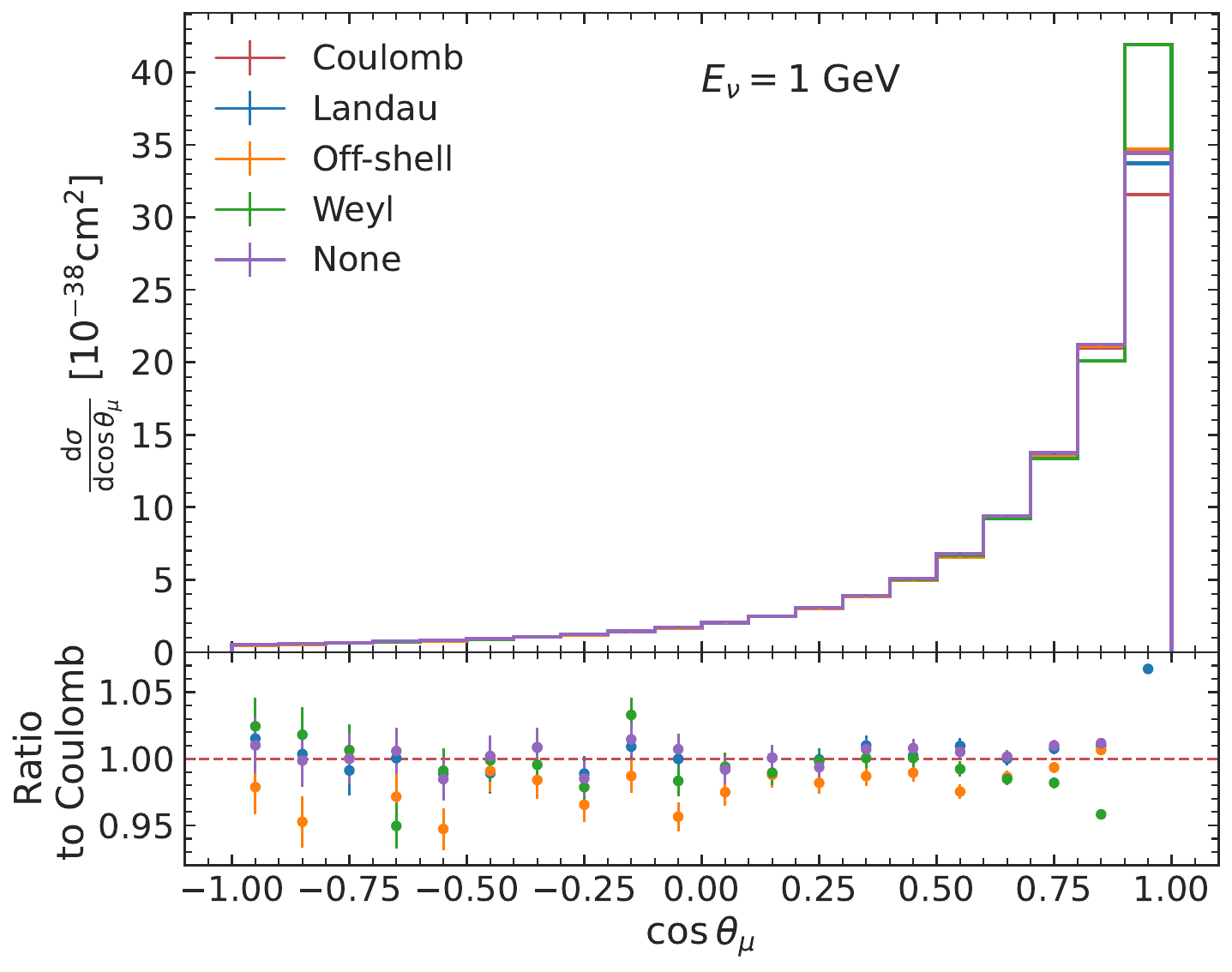}
    \\
    \includegraphics[width=\linewidth]{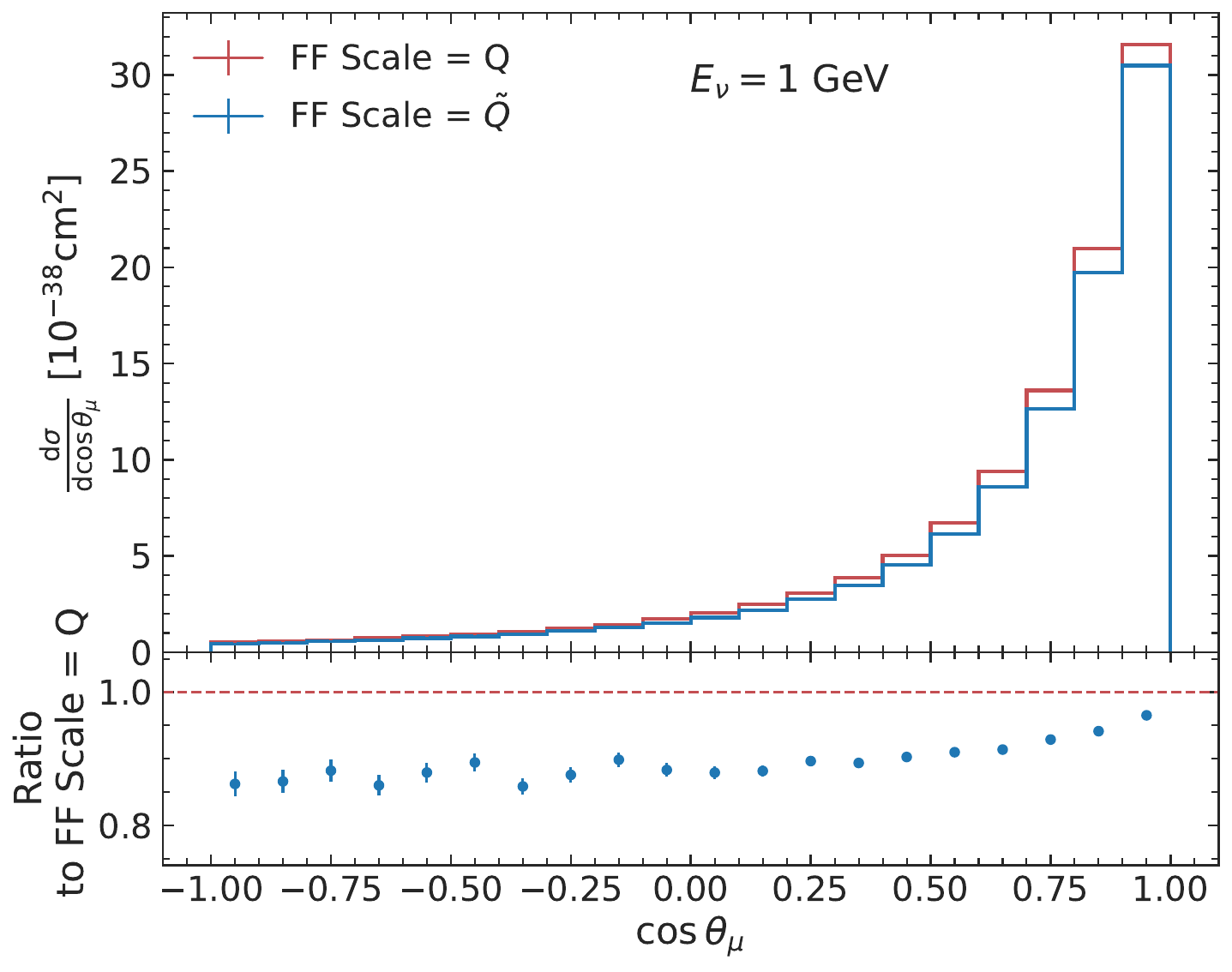}
    \caption{Same as Fig.~\ref{fig:neutrino_scattering_mom}, but for the cosine of the angle of the outgoing muon with respect to the incoming neutrino.}
    \label{fig:neutrino_scattering_cost}
\end{figure}

These different prescriptions and scale choices induce an irreducible theoretical uncertainty, when using the de Forest prescription.
A selection of results for the uncertainty from restoring current conservation is shown in Fig.~\ref{fig:electron_scattering} and Figs.~\ref{fig:neutrino_scattering_mom},~\ref{fig:neutrino_scattering_cost} for electron-nucleus and neutrino-nucleus scattering, respectively.
Events are simulated using the \textsc{Achilles} event generator~\cite{Isaacson:2022cwh}.
It is important to note that in Fig.~\ref{fig:electron_scattering}, the theory calculations only include contributions from quasielastic scattering and neglects contributions from meson-exchange currents and resonance production. Therefore, it is expected that the theory calculations under-predict the data at large energy transfer ($\omega$).
We clearly see that the difference between the Landau and Coulomb prescription and the Weyl prescription is drastic.
We can understand the major shift for the Weyl prescription by performing the calculation in the Breit frame~\cite{Streng:1979pv}.
In this frame, $\omega = 0$, and the correction factor in Eq.~\eqref{eq:weyl} diverges.
This unphysical behavior indicates that the Weyl prescription should not be considered as a method of restoring current conservation.
However, even only considering the Landau and Coulomb prescription, the numerical differences shown in Fig.~\ref{fig:electron_scattering} are on the order of 2--5\%, already above the needed precision for DUNE. 
Additionally, the ambiguity in the choice of scale for the form factors provides an additional 5--10\% uncertainty.
We remind the reader that these uncertainties are irreducible.

To address this issue, we propose a systematically improvable approach to addressing the off-shell initial state nucleon.
The insertion of the complete set of states in Eq.~\eqref{eq:SF} can be thought of as replacing a propagating nucleon from some nucleus to the hard interaction point, and then using the completeness relation to replace the propagator with a set of spinors. This technique is often used in high-energy physics to compute off-shell matrix elements which are sewn together and summed over polarization states to form complete, gauge invariant scattering amplitudes~\cite{Ballestrero:1994jn,Cachazo:2004kj,Dixon:2013uaa}.
The spinors are defined by first constructing an auxiliary spinor $w(k_0,\lambda)$ for an arbitrary massless vector $k_0$ satisfying
\begin{equation}
    w(k_0, \lambda)\bar{w}(k_0, \lambda) = \frac{1+\lambda \gamma_5}{2}\slashed{k}_0\,,
\end{equation}
with the relative phase fixed by
\begin{equation}
    w(k_0,\lambda)=\lambda\slashed{k}_1 w(k_0,-\lambda),
\end{equation}
where $k_1$ is a vector with $k_1^2=-1,\,k_0\cdot k_1=0$. With this definition of the auxiliary spinor, the spinors for a four-momentum $p$ are given as
\begin{align}
    u(p,\lambda) = \frac{\slashed{p}+m_{\rm eff}}{\sqrt{2p\cdot k_0}}w(k_0,-\lambda), \\
    v(p,\lambda)=\frac{\slashed{p}-m_{\rm eff}}{\sqrt{2p\cdot k_0}}w(k_0,-\lambda),
\end{align}
where this equation is valid for any $p^2=m_{\rm eff}^2$, even when $m_{\rm eff}$ is imaginary, or for an off-shell particle with an effective mass given from the binding energy as is the case here.
By directly using the off-shell current, we introduce a new form factor $f_3$ and an additional uncertainty associated with how the form factor scales as a function of the virtuality.
The uncertainty associated with both components can be systematically improved through new theoretical insight and experimental measurements.
As a direct consequence, there should also be a contribution to the spectral function from negative energy states. However, its effect should be small and has traditionally not been accounted for, so it will be neglected in all calculations throughout this paper. This approximation can however be improved upon in a systematic fashion.

To get a rough idea of the size of the uncertainties currently, we compare the off-shell scheme to the same set of electron and neutrino scattering plots as before in Figs.~\ref{fig:electron_scattering}, \ref{fig:neutrino_scattering_mom}, and~\ref{fig:neutrino_scattering_cost}.
The off-shell events are simulated using a custom interface between \textsc{Sherpa}~\cite{Sherpa:2024mfk} and \textsc{Achilles}, with a model defined using UFO2.0~\cite{Darme:2023jdn}.
The solid orange curve corresponds to treating the off-shell form factors as equivalent to the on-shell form factors (i.e. $f_i^{(+)}(q^2, p^2) = F_i(q^2)$, $f_3^{(\pm)}(q^2, p^2)=0$, and $f_i^{(-)}(q^2, p^2) = 0$).
The off-shell form factors can be calculated using the vector-meson dominance model with one-loop corrections~\cite{Tiemeijer:1990zp} or Chiral Perturbation Theory~\cite{Bos:1993iz}. Since these calculations are not as accurate as the lattice or experimental extraction of the on-shell form factors, the off-shell effect can be modeled by rescaling the on-shell form factors as
\begin{equation}
    f_i = \left(\frac{f_i(q^2, p^2)}{f_i(q^2, M^2)}\right)
    F_i(q^2)\,.
\end{equation}
Further efforts investigating this scheme should provide a systematically improvable prediction that will be able to provide a realistic uncertainty associated with the method at a level acceptable for DUNE and HyperK.

For the success of next-generation experiments, significant work needs to be invested into reducing the underlying theory uncertainty.
In this work, we presented a commonly neglected irreducible uncertainty associated with the restoration of gauge invariance in theory predictions.
To address this work, we propose to reconsider the evaluation of the matrix elements using off-shell initial state nucleons.
This introduces a dependency of the off-shellness of the nucleon in the form factors, and increases the required form factors from four to ten.
While the functional dependence on the off-shellness can not be directly measured experimentally or calculated currently using lattice quantum chromodynamics, it is possible to develop a model using either vector meson dominance or Chiral Perturbation theory.
Additionally, the Chiral Perturbation theory model provides a method of estimating an uncertainty associated with the truncation of higher order corrections.
Therefore, this new approach provides a systematically improvable method of providing theory predictions with a controlled prescription of obtaining theory uncertainties.
The calculation of the corrections in this scheme are left to a future work.


\begin{acknowledgments}
J.I. thanks Stefan H\"oche for helpful discussions and comments on this manuscript, Alexis Nikolakopoulos, Steven Gardiner, and Alessandro Lovato for comments on the manuscript, Noemi Rocco for explaining the de Forest description in detail and advice early on in the development of this work, Noah Steinberg for insightful discussions about the scale for the form factors.

This manuscript has been authored by Fermi Forward Discovery Group, LLC
under Contract No. 89243024CSC000002 with the U.S.\ Department of Energy,
Office of Science, Office of High Energy Physics.
The work of J.I. was supported by the U.S. Department of Energy, Office of Science, Office of Advanced Scientific Computing Research, Scientific Discovery through Advanced Computing (SciDAC-5) program, grant “NeuCol”.
\end{acknowledgments}


\bibliographystyle{apsrev4-2}
\bibliography{refs}


\end{document}